\documentclass[aps,groupedaddress,twocolumn,showpacs]{revtex4-1}
\usepackage[dvips]{graphicx}% Include figure files
\usepackage{dcolumn}% Align table columns on decimal point
\usepackage{bm}% bold math
\usepackage{amssymb}
\usepackage{epstopdf}
\usepackage{amsmath}
\begin{document}

\title{Tuning the electronic and magnetic properties of metal-doped phenanthrene by codoping method}

%\author{Jianjun Fang$^{1}$}
\author{Xun-Wang Yan$^{1}$}\email{yanxunwang@163.com}
%\author{Zhong-Yi Lu$^{1}$}\email{zlu@ruc.edu.cn}
\author{Chunfang Zhang$^{2}$}\email{zhangcf@csrc.ac.cn}
\author{Guohua Zhong$^{3}$}
\author{Jing Li$^{1}$}
\date{\today}

\affiliation{$^{1}$College of Physics and Engineering, Qufu Normal University, Qufu, Shandong 273165, China}
\affiliation{$^{2}$College of Chemistry and Environmental Science, Hebei University, Baoding, Hebei 071002}
\affiliation{$^{3}$Shenzhen Institutes of Advanced Technology, Chinese Academy of Sciences, Shenzhen 518055, China}

\begin{abstract}
  By first principles method, we have determined the geometric configuration of K/Ba-codoped phenanthrene based on the formation energy calculations, and systematically investigated its electronic and magnetic properties.
There are two bands crossing Fermi energy which mainly result from the LUMO+1 orbitals of two phenanthrene molecules in a unit cell, and the cylinder-like Fermi surface along the $\Gamma$ - Z direction reflects the two-dimension character of metallic conduction of K/Ba-codoped phenanthrene. Compared to K-doped phenanthrene, K/Ba-codoping can donate more electrons to molecule to modify the electronic structure,while the intercalation of dopants does not result in the large distortion of molecule.
  (KBa)$_1$phenanthrene is a magnetic metal with the spin moment of 0.32 $\mu_B$ per each molecule, and unexpectedly, the spins gather in one edge of molecule rather than a uniform distribution on the whole molecule.
  Our results demonstrate that codoping of monovalent and bivalent metals is an effective approach to modulate the electronic properties of metal-doped hydrocarbons.
\end{abstract}


\maketitle
\section{Introduction}
 Recently, the alkali metals intercalated aromatic molecular solids has attracted considerable interest in condensed matter physics and material science fields.
 Superconductivity in potassium-doped picene (C$_{22}$H$_{14}$) with the transition temperature of 18 K was first reported in 2010,
  \cite{Mitsuhashi2010, Artioli2014, Kubozono2016}
In subsequent experiments, phenanthrene, coronene, dibenzopentacene, chrysene, anthracene and pentacene, as well as metal dopants, were employed
 to synthesize the samples to explore the superconductivity. \cite{Wang2011, Xue2012, Kubozono2011, Okazaki2013, Artioli2015, Wu2016, Xu2017, Phan2016, Kambe2016}
 Very Recently,  Xiao-Jia Chen's group reported potassium-doped $p$-terphenyl superconductor with surprisingly high transition temperatures of 43 Kelvin and 123 Kelvin. \cite{Wang2017b}
 Rosseinsky's group and Ar\v{c}on's group synthesized the crystalline  K$_2$pentacene, K$_2$picene and triphenylide-Based K$_2$(C$_{18}$H$_{12}$)$_2$(DME), and the spin liquid state arising purely from carbon $\pi$ electrons were reported. \cite{Romero2017, Takabayashi2017, Stefancic2017}
On the other hand, there are some experimental works in past several years that do not support the appearance of superconductivity in aromatic hydrocarbons.\cite{Heguri2015}
These controversies have stimulated extensive researches on metal-doped aromatic hydrocarbon in physics and material science field.

Several kinds of metals acted as the dopants are inserted into aromatic hydrocarbons in experiment,
including alkali metal K and Rb, alkali earth metal Ba and Sr, rare earth metal La and Sm. Because of their different valence states, one metal atom can donate one, two or three electrons to molecule. The specific stoichiometries of metal atom and molecule, such as K$_3$phenenathrene (3 : 1), K$_3$picene (3 : 1), Ba$_{1.5}$phenanthrene (1.5 : 1), Sm$_1$phenanthrene(1 : 1), K$_4$coronene (4 : 1) and so on, are selected to synthesize the samples.
Theoretically, lots of important results have also been achieved. The structural and electronic properties K-doped picene with various doping levels were systematically investigated by Kosugi \textit{et al}.\cite{Kosugi2011} Verg\'es's group acquired two distinct phases of K$_3$picene after structure optimization, and recently studied the structural and electronic changes of pentacene induced by potassium doping via first principles calculations.\cite{DeAndres2011a,DeAndres2011,Guijarro2017} Tosatti's group employed the \textit{ab initio} evolutionary simulation method to search the possible structures of K$_3$picene and K$_3$phenanthrene.\cite{Naghavi2014}

The aromatic superconductor mentioned above is should be regarded as a kind of electron doping superconductor.
In experiment, the existence of multiple superconducting phases are related to different doping levels.
Besides, the donation of three electrons to one molecule is thought to be a favorable factor to superconductivity in these compounds.\cite{Mitsuhashi2010,Wang2011,Wang2011a,Huang2014,Dutta2014}
Therefore, the quantity of charge transferred to each molecule, which depends on the doping level and the dopant species, is the key factor to induce the superconductivity.
In addition, another factor mentioned here is the size of interstitial space in molecular layer.
Considering the interstitial space available in a specific aromatic crystal, only certain dopant concentrations are reasonable and feasible.
As indicated in Ref. \citenum{Yan2016}, the interstitial space in phenanthrene crystal is too small to accommodate three K atoms and the stoichiometry of K$_2$phenanthrene is the optimum ratio.

Codoping is an effective method to adjust the material electronic structures.
 The codoping of monovalent, bivalent and trivalent metals provide more choices and more possibilities, which can conveniently control the charge quantity donated to molecule with respect to single kind of metal doping.
 Although a lot of experimental and theoretical results were reported, the codoping scheme has not been mentioned in previous researches.
In this work, we take K/Ba-codoped phenanthrene for example to exhibit the tuning of their atomic structure and electronic properties by codoping method.
Our results indicate that (KBa)$_1$phenanthreneit is a good metal with a high density of states of 19.6 states/eV around Fermi energy. The Fermi surface sheet along $\Gamma$ - Z has a cylinder shape associated with the two-dimensional nature of the electronic states.
Also the unique spin density distribution on the phenanthrene molecule is found.
The codoping method we suggested provides a practical and effective approach to tune the electronic properties of the metal-doped aromatic hydrocarbons, which may be a new way to explore the superconductivity in the class of compounds.

\section{Computational details}
The plane wave basis sets and pseudopotential method were employed in our calculations. The projector augmented-wave method (PAW) pseudopotentials of C, H, K and Ba elements, with the generalized gradient approximation (GGA) for the exchange-correlation potentials, are from the subfolder C$\_s$, H$\_s$, K$\_{sv}$ and Ba$\_{sv}$ in the pseudopotential package potpaw$\_$PBE.52 supplied by Vienna Ab initio simulation package (VASP) website. \cite{PhysRevB.50.17953, PhysRevB.47.558, PhysRevB.54.11169, PhysRevLett.77.3865} The plane wave basis cutoff of 500 eV and a mesh of $8\times 8\times 6$ k-points were used for the relaxation of the lattice parameters and the internal atomic positions. The convergence thresholds of the total energy, force on atom and pressure on cell are 10$^{-5}$ eV, 0.005 eV/\AA ~and 0.1 KBar respectively. The van der Waals (vdW) interaction is included in our calculations, and the van de Waals functional we used is the vdW-DF2 scheme, proposed by Langreth and Lundqvist {\it et al}. \cite{PhysRevLett.92.246401, Roman-Perez2009, Lee2010, Klimes2011}

\section{Results and discussion}

For aromatic hydrocarbons, the molecule is composed of several benzene rings fused together, and the molecules are stacked in a herringbone pattern to form a molecular layer. Along $c$ axis, these molecular layers are aligned to construct the crystal.
 Another structural feature is that there exist a lot of interstitial space in the molecular layer and between two neighbor layers.
As shown in Fig. 1, a interstitial space in molecular layer can be treated as a hole enclosed by four molecules, and one unit cell contains two molecules and two holes in phenanthrene crystal.
When metal dopants enter into the molecular crystals, they prefer to reside in the intralayer holes instead of the region between two neighbor layers, which has been confirmed by the latest experiment reports. \cite{Romero2017,Takabayashi2017} This can be explained as follows: the interaction of dopant in a hole and $\pi$ electrons of molecule can
enhance the stability of
the doped phenanthrene system; on the other hand, the dopants in the interlayer region would be rejected by the repulsion from the H atom at the end of molecule.
Consequently,
 the insertion of the dopants into the holes in molecular layer to construct the initial geometry of unit cell is feasible and reasonable, just as most researchers did in previous studies. \cite{Kosugi2011,DeAndres2011a,DeAndres2011,Yan2016,Yan2016a,Zhong2017}.

\subsection{Atomic structure of K/Ba-codoped phenanthrene}

\begin{figure}.
\begin{center}
\includegraphics[width=7.50cm]{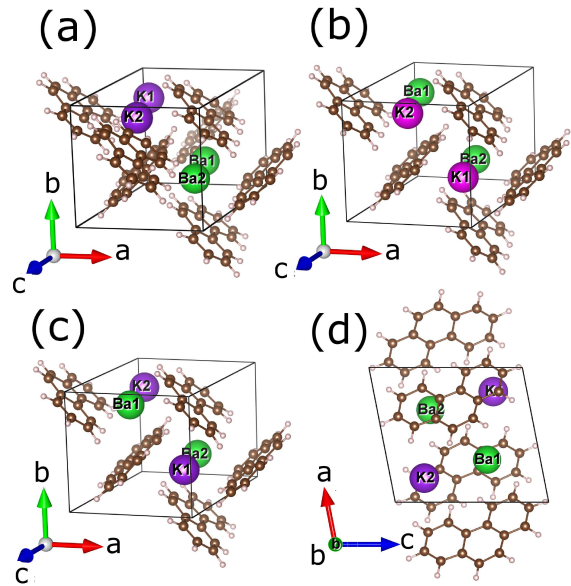}
\caption{
 The crystal structures of three structural phases of (KBa)$_1$phenanthrene. The front views along $c$ axis are shown in (a), (b) and (c) for the unit cells of (KBa)$_1$PHA-A, (KBa)$_1$PHA-B and (KBa)$_1$PHA-C, respectively, and the side views along $b$ axis for (KBa)$_1$PHA-C are shown in (d). The unit cells are all composed of two molecules, two K atoms and two Ba atoms.
    } \label{structure-1}
\end{center}
\end{figure}

The previous work has demonstrated that the optimum stoichiometry ratio of dopant and molecule is 2 : 1 for metal-doped phenanthrene crystal, which implies each hole accommodate two dopant atoms. \cite{Yan2016a}
If the two dopants are both monovalent or bivalent metal atoms, two or four electrons are provided to one phenanthrene molecule.
In order to realize the donation of three electron to a molecule, two dopants are chosen as one monovalent K atom  and one bivalent Ba atom.
A unit cell is composed of two K atoms, two Ba atoms and two phenanthrene molecules.
Therefore, three possible structures with different K/Ba dopant configurations are considered as displayed in Fig. 1 (a), (b) and (c), which are named as (KBa)$_1$PHA-A, (KBa)$_1$PHA-B and (KBa)$_1$PHA-C, respectively.
The first structure is that one hole contains two K atoms and the other hole contains two Ba atoms in the unit cell.
The second is that two different atoms sit in one hole, but two same atoms are both at the head or end of holes.
The third is that K and Ba atoms are aligned in two holes in terms of P2$_1$ space group symmetry.
After relaxation, (KBa)$_1$PHA-C is found to be most stable with the lowest energy, and the energy gain is 0.185 eV and 0.764 eV per unit cell compared to (KBa)$_1$PHA-A, (KBa)$_1$PHA-B respectively.
The optimized lattice parameters for the above three configurations are listed in Tab. 1.
We can see that the P2$_1$ group symmetry is kept in (KBa)$_1$PHA-C structure.
The $z$ coordinations are 0.75$c$ and 0.25$c$ for two K atoms, 0.33$c$ and 0.67$c$ for two Ba atoms in the unit cell of (KBa)$_1$PHA-
C structure.
Compared with K$_2$phenanthrene in Ref. \citenum{Yan2016a} , the K positions are more close to the molecular end.
In recent experiment, the measured lattice constants of K$_2$picene ($a$ = 7.64 \AA~ and $b$ = 7.93 \AA) are close to the ones of (KBa)$_1$PHA-C phase,\cite{Romero2017} indicating that the lattice parameters $a$ and $b$ in our calculations are reasonable and reliable.
\begin{table}
  \caption{The optimized lattice parameters for three structural phases of (KBa)$_1$phenanthrene.}
  \label{opti-latt}
  \begin{tabular}{lllllll}
    \hline
          phases    & a (\AA) & b (\AA) & c (\AA) &$\alpha$($^{\circ}$) &$\beta$($^{\circ}$) & $\gamma$($^{\circ}$) \\
    \hline
    (KBa)$_1$PHA-A   & 7.80 & 7.41 & 10.44 & 90.3 & 104.0 & 89.6 \\
    (KBa)$_1$PHA-B   & 7.64 & 7.58 & 10.53 & 87.7 & 105.6 & 90.3  \\
    (KBa)$_1$PHA-C   & 7.61 & 7.69 & 10.48 & 90.0 & 97.76 & 90.0 \\
    \hline
  \end{tabular}
\end{table}

\subsection{Formation energy of K/Ba-codoped phenanthrene}
In order to inspect the feasibility of the synthesis of K/Ba-codoped phenanthrene compounds, we calculate the formation energy of K/Ba-codoped phenanthrene in terms of the following expressions.

\begin{eqnarray}
%\begin{equation}
\begin{aligned}
  E_{formation}& = \frac{1}{4} \{E_{(KBa)_1PHA}-E_{PHA}-2E_{K}-2E_{Ba}\} \nonumber \\
                         %& = -0.28 ~eV \\
\end{aligned}
%\end{equation}
\end{eqnarray}

where $E_{PHA}$ and $E_{(KBa)_1PHA}$ are the energies of the unit cell of pristine phenanthrene and (KBa)$_1$phenanthrene, and
$E_{K}$ and $E_{Ba}$ are the energy of one atom in potassium and barium metal.
$\frac{1}{4}$ means that the energy is divided by the number of metal atoms in each unit cell.
For the structural phases (KBa)$_1$PHA-A, (KBa)$_1$PHA-B and (KBa)$_1$PHA-C, the formation energies are
 -0.09 eV, -0.23 eV, -0.28 eV.
(KBa)$_1$PHA-C phase has the largest formation energy, meaning that it is more easy to be synthesized in experiment than two other structural phases.
We also note that -0.28 eV per dopant atom for (KBa)$_1$PHA-C phase is a considerable large value,
 which are comparable to the formation energy of KC$_8$ (-0.285 eV per K atom), \cite{Wang2014a}
denoting that the K/Ba-codoping scheme we suggest above is feasible in the viewpoint of formation energy.

\subsection{Charge transfer and charge density distribution}
\begin{figure}.
\begin{center}
\includegraphics[width=7.50cm]{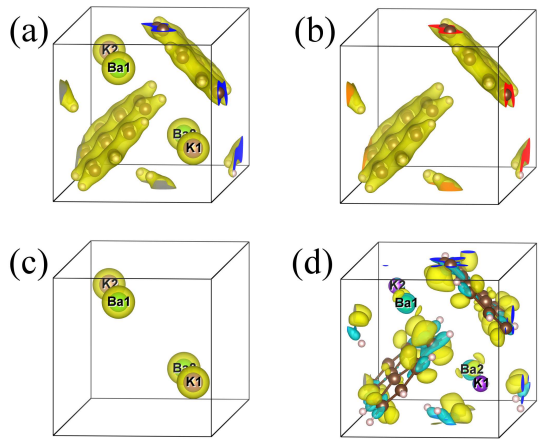}
\caption{
   (a), (b) and (c) are the charge density distribution on (KBa)$_1$PHA-C, phenanthrene and K/Ba dopant in one unit cell, and the value of charge isosurfaces is 0.172 electron/\AA$^3$. (d), the charge density difference is derived by the calculation that the total charge density of (KBa)$_1$PHA-C minus the density of parent phenanthrene and dopant K/Ba atoms respectively, and yellow and light blue isosurfaces of charge density are +0.048 and -0.048 electron/\AA$^3$, corresponding to the electron accumulation and depletion respectively.} \label{charge-diff}
\end{center}
\end{figure}

Charge transfer plays an important role on tuning the electronic properties and inducing the superconductivity in metal-doped aromatic solids.
Especially, the donation of three electrons to each molecule in metal-doped polycyclic aromatic hydrocarbons are regarded as a fundamental factor to superconductivity emergence in experiment,\cite{Mitsuhashi2010,Wang2011,Wang2011a,Huang2014} and this assumption is supported by the subsequent theoretical results that molecule valence 3 is special in the class of metal-doped organic systems.\cite{Dutta2014}
But on the basis of theoretical studies of K$_x$phenanthrene, there is not enough space to accommodate three potassium atoms in each interstitial room enclosed by four molecule, namely K$_3$phenanthrene is impossible to be synthesized from the viewpoint of formation energy. \cite{Yan2016}
Hence, codoping is a possible scheme to realize three electrons transferring to each molecule.
For (KBa)$_1$PHA-C phase, we inspect the charge transfer from K and Ba dopant to phenanthrene molecule, and find that K and Ba atom donate 0.82 and 1.36 electron according to the Bader charge analysis. \cite{Henkelman2006}
The C atoms at the edge of molecule, which are close to the metal dopants, attract more charge around them with the maximum charge gain of 0.19 electron on C atom.
Fig. \ref{charge-diff} displays the charge density distribution of (KBa)$_1$PHA-C, phenanthrene and K/Ba dopant in one unit cell and the charge density difference which is obtained by subtracting the charge density of phenanthrene and K/Ba dopant from the total charge density of (KBa)$_1$PHA-C.
In Fig. \ref{charge-diff} (d), yellow and light blue isosurfaces of charge density means the electron accumulation and depletion respectively, with the values of +0.048 and -0.048 electron/\AA$^3$.
As can be seen, metal atom doping can result in the charge accumulation of the $\pi$ bond of molecule, while an unexpected phenomena is that the charge of C-H bond is reduced.

\subsection{Magnetic properties}
In metal-doped aromatic hydrocarbons, the mechanism of the superconducting pair is in debate,
but there exist some hints to manifest the unconventional nature of superconductivity, such as the obvious positive pressure dependence of T$_c$ and the existence of local magnetic moment in K-, Ba-, Sr-, La-, and Sm-doped phenanthrene superconductors.\cite{Wang2011,Wang2011a, Wang2012}
The reported magnetic susceptibility measurement showed 0.22 $\mu_B$ per molecule in the K-doped phenanthrene, \cite{Wang2011} 0.5 $\mu_B$ per molecule in K-doped anthracene, \cite{Phan2014} and local spin in K-doped picene was also detected by electron spin-resonance spectroscopy. \cite{Mitsuhashi2010}
These results indicate that the charge transfer from metal dopant to molecule can induce local spins in metal-doped aromatic hydrocarbons.

We performed the spin-polarized calculations to examine the magnetic properties of K/Ba-codoped phenanthrene with the lowest energy phase (KBa)$_1$PHA-C. All spins of atoms in unit cell are aligned parallel to construct the ferromagnetic ordering, and the spins in a molecule is parallel and the spins are anti-parallel between two molecules to form the antiferromagnetic ordering.
The energy of unit cell in antiferromagnetic ordering is 3.7 meV lower than the energy in nonmagnetic state, and 0.6 meV lower than ferromagnetic state.
These results indicate that K/Ba-doped phenanthrene has a magnetic ground state, and antiferromagnetism is more favorable.
The local spin per molecule is about 0.32 $\mu_B$ in antiferromagnetic state,
which is in a good agreement with previous experimental and theoretical reports. \cite{Phan2014,Zhong2014,Yan2016a}

\begin{figure}.
\includegraphics[width=7.50cm]{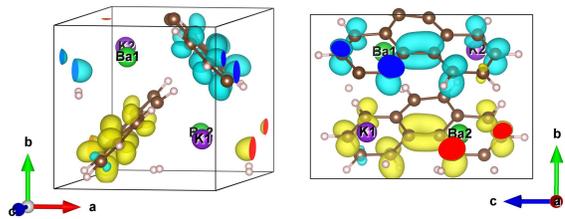}
\caption{Calculated spin density of (KBa)$_1$PHA-C in antiferromagnetic state. The two colors of electronic cloud correspond to two spin directions. Isosurface value is 0.002 electron/\AA$^3$. The left and right panels are related to the front view and side view.} \label{SPIN}
\end{figure}

Fig. \ref{SPIN} displays the spin density of (KBa)$_1$PHA-C in antiferromagnetic state, which is the difference of charge density between spin up charge and spin down charge.
The local spin concentrates upon the one of two edges of molecule rather than a uniform distribution on whole molecule.
 Also, the spin density distribution is not consistent with the charge difference distribution in Fig. \ref{charge-diff} (d), where the additional charge coming from the dopants reside on two edges of molecule. This difference of spin density and additional charge distributions is more obvious on the middle benzene ring of phenanthrene molecule.

\subsection{Electronic properties of K/Ba-codoped phenanthrene}
\begin{figure}.
\includegraphics[width=7.50cm]{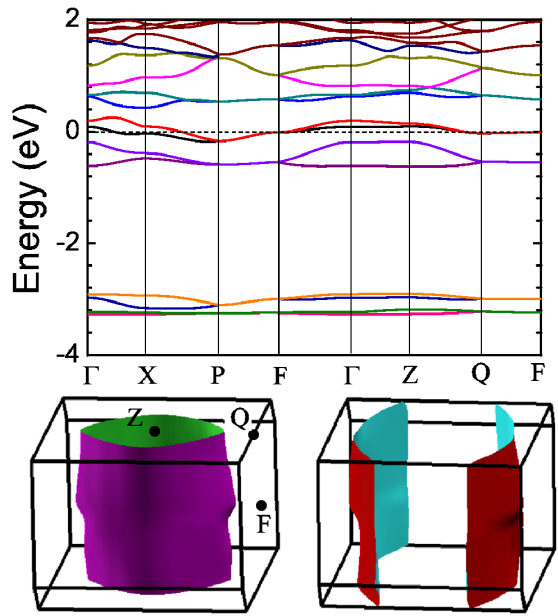}
\caption{Upper panel: Energy band structure of (KBa)$_1$PHA-C in nonmagnetic state with two bands crossing the Fermi level; Lower panels: Fermi surfaces for the two bands. The Fermi energy is set to zero.
} \label{band-fs}
\end{figure}

We choose (KBa)$_1$PHA-C phase to exhibit the electronic properties of K/Ba-codoped phenanthrene, and
the band structure of (KBa)$_1$PHA-C in nonmagnetic state is shown in upper panel of Fig. \ref{band-fs}. One can see that (KBa)$_1$PHA-C is a metal with two energy bands crossing the Fermi level.
The feature of band structure is that two bands form a band pair, which comes from two equivalent orbitals belonging to two molecules in a unit cell.
Because of the nominal three electrons transferred to each molecule in (KBa)$_1$phenanthrene, two of them occupy the LUMO (lowest unoccupied molecular orbital) and the third electron fills the LUMO + 1 orbital,
which leads to the band pair from LUMO + 1 orbitals being half-filled and crossing the Fermi level.

Fig. \ref{band-fs} lower panel shows the Fermi surface of (KBa)$_1$PHA-C in nonmagnetic state.
The first sheet of the surface has a cylinder-like shape along the $\Gamma$ - Z direction. So this Fermi surface obviously reflects the two-dimension character of metallic conduction of the molecular layer parallel to $ab$ crystal plane. The second surfaces are large arched sheets and partly take on the two-dimensional character of the electronic states of the molecular layer.

\begin{figure}.
\includegraphics[width=7.50cm]{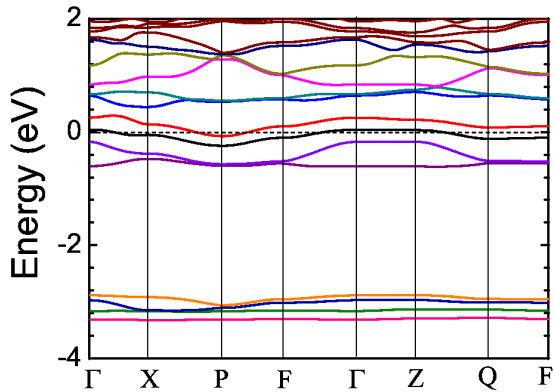}
\caption{Energy band structure of (KBa)$_1$PHA-C in antiferromagnetic state. The two bands crossing Fermi level have less overlap with respect to the ones in nonmagnetic case.
} \label{band-AFM}
\end{figure}

We also perform the spin-polarized calculation of (KBa)$_1$PHA-C phase by assigning a finite magnetic moment on each atom.
The energy bands in antiferromagnetic state are shown in Fig. \ref{band-AFM}~. We find that two bands of a band pair at Fermi level are separated by the the spin-polarization effect, which have a obvious change with respect to the related two bands in nonmagnetic state in Fig. \ref{band-fs}.
Meanwhile, the total density of states (DOS) at Fermi level have been reduced due to spin polarization.
In our calculations, the value of total DOS in antiferromagnetic state is 11.1 states/eV per unit cell, lower that the value of 19.6 states/eV per unit cell in nonmagnetic state.

\section{Discussions and Conclusions}
Codoping with different kinds of metal atoms can exert the obvious influence on the electronic and magnetic characters of parent compounds.
If we only choose K or Ba as dopant in metal doped phenanthrene, (KK)$_1$PHA or (BaBa)$_1$PHA phase will be generated, in which two or four electrons are provided to each molecule. The band pair from two LUMO orbitals or two LUMO + 1 orbital is just fully occupied in the rigid band approximation, rather than partly occupied. If the trivalent La (or Sm) and bivalent Ba atoms are used as dopant in (BaLa)$_1$phenanthrene crystal, five electrons are provided for one molecule and the bands crossing Fermi level are from the LUMO + 2 orbitals of phenanthrene molecule.
As for metal-doped picene and other suitable aromatic hydrocarbons, the interstitial space in molecule layer is large enough to accommodate three or more metal atoms per molecule.
The combinations of monovalent, bivalent and trivalent metal atoms have more possibilities to control the charge transfer to molecule and influence their atomic structures.
Therefore, codoping of two or more kinds of metal atoms with +1, +2 and +3 valence can effectively modulate the electronic structure for exploring the promising superconductivity in metal-doped aromatic hydrocarbons.

In summary, we have investigated the atomic structure, magnetic and electronic properties of K/Ba-codoped phenanthrene with first principle simulation method. The formation energy of (KBa)$_1$PHA-C phase is -0.28 eV per metal atom, which is such a large value that implies the process of K/Ba-codoping being feasible in energetic viewpoint.
The additional charge donated by metal atoms mainly distributed on the two edges of molecule, but the local spin reside on one edge of molecule.
The detailed spin distribution and its uniqueness is reported for the first time in this class of metal-doped aromatic hydrocarbons.
(K/Ba)-codoped phenanthrene with (KBa)$_1$PHA-C phase has the antiferromagnetic ground state with the magnetic moment of 0.32 $\mu_B$ per molecule, which is consistent with the previous experimental results in related systems.
%The calculational results indicate that
 K/Ba-codoping can tune effectively the electronic structure of metal-doped phenanthrene, in which three electrons are donated to each phenanthrene molecule to ensure the pair of bands from LUMO + 1 orbitals crossing the Fermi level.
Thus, our work provides an effective method to tune the electronic properties, and suggests that codoping is a promising routine to synthesize the exotic electronic materials including the aromatic superconductors and carbon-based spin liquid materials.

\begin{acknowledgments}
This work was supported by the National Natural Science Foundation of China under Grants
Nos. 11474004, 11274335, U1204108, 11604179, 11704024, and Special Program for Applied Research on Super Computation of the NSFC-Guangdong Joint Fund (the second phase) under Grant No.U1501501.
\end{acknowledgments}

\bibliography{KBa-codoping}

\begin{thebibliography}{41}%
\makeatletter
\providecommand \@ifxundefined [1]{%
 \@ifx{#1\undefined}
}%
\providecommand \@ifnum [1]{%
 \ifnum #1\expandafter \@firstoftwo
 \else \expandafter \@secondoftwo
 \fi
}%
\providecommand \@ifx [1]{%
 \ifx #1\expandafter \@firstoftwo
 \else \expandafter \@secondoftwo
 \fi
}%
\providecommand \natexlab [1]{#1}%
\providecommand \enquote  [1]{``#1''}%
\providecommand \bibnamefont  [1]{#1}%
\providecommand \bibfnamefont [1]{#1}%
\providecommand \citenamefont [1]{#1}%
\providecommand \href@noop [0]{\@secondoftwo}%
\providecommand \href [0]{\begingroup \@sanitize@url \@href}%
\providecommand \@href[1]{\@@startlink{#1}\@@href}%
\providecommand \@@href[1]{\endgroup#1\@@endlink}%
\providecommand \@sanitize@url [0]{\catcode `\\12\catcode `\$12\catcode
  `\&12\catcode `\#12\catcode `\^12\catcode `\_12\catcode `\%12\relax}%
\providecommand \@@startlink[1]{}%
\providecommand \@@endlink[0]{}%
\providecommand \url  [0]{\begingroup\@sanitize@url \@url }%
\providecommand \@url [1]{\endgroup\@href {#1}{\urlprefix }}%
\providecommand \urlprefix  [0]{URL }%
\providecommand \Eprint [0]{\href }%
\providecommand \doibase [0]{http://dx.doi.org/}%
\providecommand \selectlanguage [0]{\@gobble}%
\providecommand \bibinfo  [0]{\@secondoftwo}%
\providecommand \bibfield  [0]{\@secondoftwo}%
\providecommand \translation [1]{[#1]}%
\providecommand \BibitemOpen [0]{}%
\providecommand \bibitemStop [0]{}%
\providecommand \bibitemNoStop [0]{.\EOS\space}%
\providecommand \EOS [0]{\spacefactor3000\relax}%
\providecommand \BibitemShut  [1]{\csname bibitem#1\endcsname}%
\let\auto@bib@innerbib\@empty
%</preamble>
\bibitem [{\citenamefont {Mitsuhashi}\ \emph {et~al.}(2010)\citenamefont
  {Mitsuhashi}, \citenamefont {Suzuki}, \citenamefont {Yamanari}, \citenamefont
  {Mitamura}, \citenamefont {Kambe}, \citenamefont {Ikeda}, \citenamefont
  {Okamoto}, \citenamefont {Fujiwara}, \citenamefont {Yamaji}, \citenamefont
  {Kawasaki}, \citenamefont {Maniwa},\ and\ \citenamefont
  {Kubozono}}]{Mitsuhashi2010}%
  \BibitemOpen
  \bibfield  {author} {\bibinfo {author} {\bibfnamefont {R.}~\bibnamefont
  {Mitsuhashi}}, \bibinfo {author} {\bibfnamefont {Y.}~\bibnamefont {Suzuki}},
  \bibinfo {author} {\bibfnamefont {Y.}~\bibnamefont {Yamanari}}, \bibinfo
  {author} {\bibfnamefont {H.}~\bibnamefont {Mitamura}}, \bibinfo {author}
  {\bibfnamefont {T.}~\bibnamefont {Kambe}}, \bibinfo {author} {\bibfnamefont
  {N.}~\bibnamefont {Ikeda}}, \bibinfo {author} {\bibfnamefont
  {H.}~\bibnamefont {Okamoto}}, \bibinfo {author} {\bibfnamefont
  {A.}~\bibnamefont {Fujiwara}}, \bibinfo {author} {\bibfnamefont
  {M.}~\bibnamefont {Yamaji}}, \bibinfo {author} {\bibfnamefont
  {N.}~\bibnamefont {Kawasaki}}, \bibinfo {author} {\bibfnamefont
  {Y.}~\bibnamefont {Maniwa}}, \ and\ \bibinfo {author} {\bibfnamefont
  {Y.}~\bibnamefont {Kubozono}},\ }\href {\doibase 10.1038/nature08859}
  {\bibfield  {journal} {\bibinfo  {journal} {Nature}\ }\textbf {\bibinfo
  {volume} {464}},\ \bibinfo {pages} {76} (\bibinfo {year} {2010})}\BibitemShut
  {NoStop}%
\bibitem [{\citenamefont {Artioli}\ and\ \citenamefont
  {Malavasi}(2014)}]{Artioli2014}%
  \BibitemOpen
  \bibfield  {author} {\bibinfo {author} {\bibfnamefont {G.~A.}\ \bibnamefont
  {Artioli}}\ and\ \bibinfo {author} {\bibfnamefont {L.}~\bibnamefont
  {Malavasi}},\ }\href {\doibase 10.1039/c3tc32326a} {\bibfield  {journal}
  {\bibinfo  {journal} {Journal of Materials Chemistry C}\ }\textbf {\bibinfo
  {volume} {2}},\ \bibinfo {pages} {1577} (\bibinfo {year} {2014})}\BibitemShut
  {NoStop}%
\bibitem [{\citenamefont {Kubozono}\ \emph {et~al.}(2016)\citenamefont
  {Kubozono}, \citenamefont {Eguchi}, \citenamefont {Goto}, \citenamefont
  {Hamao}, \citenamefont {Kambe}, \citenamefont {Terao}, \citenamefont
  {Nishiyama}, \citenamefont {Zheng}, \citenamefont {Miao},\ and\ \citenamefont
  {Okamoto}}]{Kubozono2016}%
  \BibitemOpen
  \bibfield  {author} {\bibinfo {author} {\bibfnamefont {Y.}~\bibnamefont
  {Kubozono}}, \bibinfo {author} {\bibfnamefont {R.}~\bibnamefont {Eguchi}},
  \bibinfo {author} {\bibfnamefont {H.}~\bibnamefont {Goto}}, \bibinfo {author}
  {\bibfnamefont {S.}~\bibnamefont {Hamao}}, \bibinfo {author} {\bibfnamefont
  {T.}~\bibnamefont {Kambe}}, \bibinfo {author} {\bibfnamefont
  {T.}~\bibnamefont {Terao}}, \bibinfo {author} {\bibfnamefont
  {S.}~\bibnamefont {Nishiyama}}, \bibinfo {author} {\bibfnamefont
  {L.}~\bibnamefont {Zheng}}, \bibinfo {author} {\bibfnamefont
  {X.}~\bibnamefont {Miao}}, \ and\ \bibinfo {author} {\bibfnamefont
  {H.}~\bibnamefont {Okamoto}},\ }\href {\doibase
  10.1088/0953-8984/28/33/334001} {\bibfield  {journal} {\bibinfo  {journal}
  {Journal of Physics: Condensed Matter}\ }\textbf {\bibinfo {volume} {28}},\
  \bibinfo {pages} {334001} (\bibinfo {year} {2016})}\BibitemShut {NoStop}%
\bibitem [{\citenamefont {Wang}\ \emph
  {et~al.}(2011{\natexlab{a}})\citenamefont {Wang}, \citenamefont {Liu},
  \citenamefont {Gui}, \citenamefont {Xie}, \citenamefont {Yan}, \citenamefont
  {Ying}, \citenamefont {Luo},\ and\ \citenamefont {Chen}}]{Wang2011}%
  \BibitemOpen
  \bibfield  {author} {\bibinfo {author} {\bibfnamefont {X.~F.}\ \bibnamefont
  {Wang}}, \bibinfo {author} {\bibfnamefont {R.~H.}\ \bibnamefont {Liu}},
  \bibinfo {author} {\bibfnamefont {Z.}~\bibnamefont {Gui}}, \bibinfo {author}
  {\bibfnamefont {Y.~L.}\ \bibnamefont {Xie}}, \bibinfo {author} {\bibfnamefont
  {Y.~J.}\ \bibnamefont {Yan}}, \bibinfo {author} {\bibfnamefont {J.~J.}\
  \bibnamefont {Ying}}, \bibinfo {author} {\bibfnamefont {X.~G.}\ \bibnamefont
  {Luo}}, \ and\ \bibinfo {author} {\bibfnamefont {X.~H.}\ \bibnamefont
  {Chen}},\ }\href {\doibase 10.1038/ncomms1513} {\bibfield  {journal}
  {\bibinfo  {journal} {Nature communications}\ }\textbf {\bibinfo {volume}
  {2}},\ \bibinfo {pages} {507} (\bibinfo {year}
  {2011}{\natexlab{a}})}\BibitemShut {NoStop}%
\bibitem [{\citenamefont {Xue}\ \emph {et~al.}(2012)\citenamefont {Xue},
  \citenamefont {Cao}, \citenamefont {Wang}, \citenamefont {Wu}, \citenamefont
  {Yang}, \citenamefont {Dong}, \citenamefont {He}, \citenamefont {Li},\ and\
  \citenamefont {Chen}}]{Xue2012}%
  \BibitemOpen
  \bibfield  {author} {\bibinfo {author} {\bibfnamefont {M.}~\bibnamefont
  {Xue}}, \bibinfo {author} {\bibfnamefont {T.}~\bibnamefont {Cao}}, \bibinfo
  {author} {\bibfnamefont {D.}~\bibnamefont {Wang}}, \bibinfo {author}
  {\bibfnamefont {Y.}~\bibnamefont {Wu}}, \bibinfo {author} {\bibfnamefont
  {H.}~\bibnamefont {Yang}}, \bibinfo {author} {\bibfnamefont {X.}~\bibnamefont
  {Dong}}, \bibinfo {author} {\bibfnamefont {J.}~\bibnamefont {He}}, \bibinfo
  {author} {\bibfnamefont {F.}~\bibnamefont {Li}}, \ and\ \bibinfo {author}
  {\bibfnamefont {G.~F.}\ \bibnamefont {Chen}},\ }\href {\doibase
  10.1038/srep00389} {\bibfield  {journal} {\bibinfo  {journal} {Scientific
  reports}\ }\textbf {\bibinfo {volume} {2}},\ \bibinfo {pages} {389} (\bibinfo
  {year} {2012})}\BibitemShut {NoStop}%
\bibitem [{\citenamefont {Kubozono}\ \emph {et~al.}(2011)\citenamefont
  {Kubozono}, \citenamefont {Mitamura}, \citenamefont {Lee}, \citenamefont
  {He}, \citenamefont {Yamanari}, \citenamefont {Takahashi}, \citenamefont
  {Suzuki}, \citenamefont {Kaji}, \citenamefont {Eguchi}, \citenamefont
  {Akaike}, \citenamefont {Kambe}, \citenamefont {Okamoto}, \citenamefont
  {Fujiwara}, \citenamefont {Kato}, \citenamefont {Kosugi},\ and\ \citenamefont
  {Aoki}}]{Kubozono2011}%
  \BibitemOpen
  \bibfield  {author} {\bibinfo {author} {\bibfnamefont {Y.}~\bibnamefont
  {Kubozono}}, \bibinfo {author} {\bibfnamefont {H.}~\bibnamefont {Mitamura}},
  \bibinfo {author} {\bibfnamefont {X.}~\bibnamefont {Lee}}, \bibinfo {author}
  {\bibfnamefont {X.}~\bibnamefont {He}}, \bibinfo {author} {\bibfnamefont
  {Y.}~\bibnamefont {Yamanari}}, \bibinfo {author} {\bibfnamefont
  {Y.}~\bibnamefont {Takahashi}}, \bibinfo {author} {\bibfnamefont
  {Y.}~\bibnamefont {Suzuki}}, \bibinfo {author} {\bibfnamefont
  {Y.}~\bibnamefont {Kaji}}, \bibinfo {author} {\bibfnamefont {R.}~\bibnamefont
  {Eguchi}}, \bibinfo {author} {\bibfnamefont {K.}~\bibnamefont {Akaike}},
  \bibinfo {author} {\bibfnamefont {T.}~\bibnamefont {Kambe}}, \bibinfo
  {author} {\bibfnamefont {H.}~\bibnamefont {Okamoto}}, \bibinfo {author}
  {\bibfnamefont {A.}~\bibnamefont {Fujiwara}}, \bibinfo {author}
  {\bibfnamefont {T.}~\bibnamefont {Kato}}, \bibinfo {author} {\bibfnamefont
  {T.}~\bibnamefont {Kosugi}}, \ and\ \bibinfo {author} {\bibfnamefont
  {H.}~\bibnamefont {Aoki}},\ }\href {\doibase 10.1039/c1cp20961b} {\bibfield
  {journal} {\bibinfo  {journal} {Physical chemistry chemical physics : PCCP}\
  }\textbf {\bibinfo {volume} {13}},\ \bibinfo {pages} {16476} (\bibinfo {year}
  {2011})}\BibitemShut {NoStop}%
\bibitem [{\citenamefont {Okazaki}\ \emph {et~al.}(2013)\citenamefont
  {Okazaki}, \citenamefont {Jabuchi}, \citenamefont {Wakita}, \citenamefont
  {Kato}, \citenamefont {Muraoka},\ and\ \citenamefont {Yokoya}}]{Okazaki2013}%
  \BibitemOpen
  \bibfield  {author} {\bibinfo {author} {\bibfnamefont {H.}~\bibnamefont
  {Okazaki}}, \bibinfo {author} {\bibfnamefont {T.}~\bibnamefont {Jabuchi}},
  \bibinfo {author} {\bibfnamefont {T.}~\bibnamefont {Wakita}}, \bibinfo
  {author} {\bibfnamefont {T.}~\bibnamefont {Kato}}, \bibinfo {author}
  {\bibfnamefont {Y.}~\bibnamefont {Muraoka}}, \ and\ \bibinfo {author}
  {\bibfnamefont {T.}~\bibnamefont {Yokoya}},\ }\href {\doibase
  10.1103/PhysRevB.88.245414} {\bibfield  {journal} {\bibinfo  {journal}
  {Physical Review B}\ }\textbf {\bibinfo {volume} {88}},\ \bibinfo {pages}
  {245414} (\bibinfo {year} {2013})}\BibitemShut {NoStop}%
\bibitem [{\citenamefont {Artioli}\ \emph {et~al.}(2015)\citenamefont
  {Artioli}, \citenamefont {Hammerath}, \citenamefont {Mozzati}, \citenamefont
  {Carretta}, \citenamefont {Corana}, \citenamefont {Mannucci}, \citenamefont
  {Margadonna},\ and\ \citenamefont {Malavasi}}]{Artioli2015}%
  \BibitemOpen
  \bibfield  {author} {\bibinfo {author} {\bibfnamefont {G.~A.}\ \bibnamefont
  {Artioli}}, \bibinfo {author} {\bibfnamefont {F.}~\bibnamefont {Hammerath}},
  \bibinfo {author} {\bibfnamefont {M.~C.}\ \bibnamefont {Mozzati}}, \bibinfo
  {author} {\bibfnamefont {P.}~\bibnamefont {Carretta}}, \bibinfo {author}
  {\bibfnamefont {F.}~\bibnamefont {Corana}}, \bibinfo {author} {\bibfnamefont
  {B.}~\bibnamefont {Mannucci}}, \bibinfo {author} {\bibfnamefont
  {S.}~\bibnamefont {Margadonna}}, \ and\ \bibinfo {author} {\bibfnamefont
  {L.}~\bibnamefont {Malavasi}},\ }\href {\doibase 10.1039/C4CC07879A}
  {\bibfield  {journal} {\bibinfo  {journal} {Chem. Commun.}\ }\textbf
  {\bibinfo {volume} {51}},\ \bibinfo {pages} {1092} (\bibinfo {year}
  {2015})}\BibitemShut {NoStop}%
\bibitem [{\citenamefont {Wu}\ \emph {et~al.}(2016)\citenamefont {Wu},
  \citenamefont {Xu}, \citenamefont {Wang},\ and\ \citenamefont
  {Xiao}}]{Wu2016}%
  \BibitemOpen
  \bibfield  {author} {\bibinfo {author} {\bibfnamefont {X.}~\bibnamefont
  {Wu}}, \bibinfo {author} {\bibfnamefont {C.}~\bibnamefont {Xu}}, \bibinfo
  {author} {\bibfnamefont {K.}~\bibnamefont {Wang}}, \ and\ \bibinfo {author}
  {\bibfnamefont {X.}~\bibnamefont {Xiao}},\ }\href {\doibase
  10.1021/acs.jpcc.6b03686} {\bibfield  {journal} {\bibinfo  {journal} {The
  Journal of Physical Chemistry C}\ }\textbf {\bibinfo {volume} {120}},\
  \bibinfo {pages} {15446} (\bibinfo {year} {2016})}\BibitemShut {NoStop}%
\bibitem [{\citenamefont {Xu}\ \emph {et~al.}(2017)\citenamefont {Xu},
  \citenamefont {Que}, \citenamefont {Zhuang}, \citenamefont {Lin},
  \citenamefont {Wu}, \citenamefont {Wang},\ and\ \citenamefont
  {Xiao}}]{Xu2017}%
  \BibitemOpen
  \bibfield  {author} {\bibinfo {author} {\bibfnamefont {C.}~\bibnamefont
  {Xu}}, \bibinfo {author} {\bibfnamefont {Y.}~\bibnamefont {Que}}, \bibinfo
  {author} {\bibfnamefont {Y.}~\bibnamefont {Zhuang}}, \bibinfo {author}
  {\bibfnamefont {Z.}~\bibnamefont {Lin}}, \bibinfo {author} {\bibfnamefont
  {X.}~\bibnamefont {Wu}}, \bibinfo {author} {\bibfnamefont {K.}~\bibnamefont
  {Wang}}, \ and\ \bibinfo {author} {\bibfnamefont {X.}~\bibnamefont {Xiao}},\
  }\href {\doibase 10.1021/acs.jpcb.7b05140} {\bibfield  {journal} {\bibinfo
  {journal} {The Journal of Physical Chemistry B}\ ,\ \bibinfo {pages}
  {acs.jpcb.7b05140}} (\bibinfo {year} {2017})}\BibitemShut {NoStop}%
\bibitem [{\citenamefont {Phan}\ \emph {et~al.}(2016)\citenamefont {Phan},
  \citenamefont {Heguri}, \citenamefont {Tamura}, \citenamefont {Nakano},
  \citenamefont {Nozue},\ and\ \citenamefont {Tanigaki}}]{Phan2016}%
  \BibitemOpen
  \bibfield  {author} {\bibinfo {author} {\bibfnamefont {Q.~T.~N.}\
  \bibnamefont {Phan}}, \bibinfo {author} {\bibfnamefont {S.}~\bibnamefont
  {Heguri}}, \bibinfo {author} {\bibfnamefont {H.}~\bibnamefont {Tamura}},
  \bibinfo {author} {\bibfnamefont {T.}~\bibnamefont {Nakano}}, \bibinfo
  {author} {\bibfnamefont {Y.}~\bibnamefont {Nozue}}, \ and\ \bibinfo {author}
  {\bibfnamefont {K.}~\bibnamefont {Tanigaki}},\ }\href {\doibase
  10.1103/PhysRevB.93.075130} {\bibfield  {journal} {\bibinfo  {journal}
  {Physical Review B}\ }\textbf {\bibinfo {volume} {93}},\ \bibinfo {pages}
  {075130} (\bibinfo {year} {2016})}\BibitemShut {NoStop}%
\bibitem [{\citenamefont {Kambe}\ \emph {et~al.}(2016)\citenamefont {Kambe},
  \citenamefont {Nishiyama}, \citenamefont {Nguyen}, \citenamefont {Terao},
  \citenamefont {Izumi}, \citenamefont {Sakai}, \citenamefont {Zheng},
  \citenamefont {Goto}, \citenamefont {Itoh}, \citenamefont {Onji},
  \citenamefont {Kobayashi}, \citenamefont {Sugino}, \citenamefont {Gohda},
  \citenamefont {Okamoto},\ and\ \citenamefont {Kubozono}}]{Kambe2016}%
  \BibitemOpen
  \bibfield  {author} {\bibinfo {author} {\bibfnamefont {T.}~\bibnamefont
  {Kambe}}, \bibinfo {author} {\bibfnamefont {S.}~\bibnamefont {Nishiyama}},
  \bibinfo {author} {\bibfnamefont {H.~L.~T.}\ \bibnamefont {Nguyen}}, \bibinfo
  {author} {\bibfnamefont {T.}~\bibnamefont {Terao}}, \bibinfo {author}
  {\bibfnamefont {M.}~\bibnamefont {Izumi}}, \bibinfo {author} {\bibfnamefont
  {Y.}~\bibnamefont {Sakai}}, \bibinfo {author} {\bibfnamefont
  {L.}~\bibnamefont {Zheng}}, \bibinfo {author} {\bibfnamefont
  {H.}~\bibnamefont {Goto}}, \bibinfo {author} {\bibfnamefont {Y.}~\bibnamefont
  {Itoh}}, \bibinfo {author} {\bibfnamefont {T.}~\bibnamefont {Onji}}, \bibinfo
  {author} {\bibfnamefont {T.~C.}\ \bibnamefont {Kobayashi}}, \bibinfo {author}
  {\bibfnamefont {H.}~\bibnamefont {Sugino}}, \bibinfo {author} {\bibfnamefont
  {S.}~\bibnamefont {Gohda}}, \bibinfo {author} {\bibfnamefont
  {H.}~\bibnamefont {Okamoto}}, \ and\ \bibinfo {author} {\bibfnamefont
  {Y.}~\bibnamefont {Kubozono}},\ }\href {\doibase
  10.1088/0953-8984/28/44/444001} {\bibfield  {journal} {\bibinfo  {journal}
  {Journal of Physics: Condensed Matter}\ }\textbf {\bibinfo {volume} {28}},\
  \bibinfo {pages} {444001} (\bibinfo {year} {2016})}\BibitemShut {NoStop}%
\bibitem [{\citenamefont {Wang}\ \emph {et~al.}(2017)\citenamefont {Wang},
  \citenamefont {Gao}, \citenamefont {Huang},\ and\ \citenamefont
  {Chen}}]{Wang2017b}%
  \BibitemOpen
  \bibfield  {author} {\bibinfo {author} {\bibfnamefont {R.-S.}\ \bibnamefont
  {Wang}}, \bibinfo {author} {\bibfnamefont {Y.}~\bibnamefont {Gao}}, \bibinfo
  {author} {\bibfnamefont {Z.-B.}\ \bibnamefont {Huang}}, \ and\ \bibinfo
  {author} {\bibfnamefont {X.-J.}\ \bibnamefont {Chen}},\ }\href
  {http://arxiv.org/abs/1703.05804} {\bibfield  {journal} {\bibinfo  {journal}
  {arXiv1703.05804}\ } (\bibinfo {year} {2017})},\ \Eprint
  {http://arxiv.org/abs/1703.05804} {arXiv:1703.05804} \BibitemShut {NoStop}%
\bibitem [{\citenamefont {Romero}\ \emph {et~al.}(2017)\citenamefont {Romero},
  \citenamefont {Pitcher}, \citenamefont {Hiley}, \citenamefont {Whitehead},
  \citenamefont {Kar}, \citenamefont {Ganin}, \citenamefont {Antypov},
  \citenamefont {Collins}, \citenamefont {Dyer}, \citenamefont {Klupp},
  \citenamefont {Colman}, \citenamefont {Prassides},\ and\ \citenamefont
  {Rosseinsky}}]{Romero2017}%
  \BibitemOpen
  \bibfield  {author} {\bibinfo {author} {\bibfnamefont {F.~D.}\ \bibnamefont
  {Romero}}, \bibinfo {author} {\bibfnamefont {M.~J.}\ \bibnamefont {Pitcher}},
  \bibinfo {author} {\bibfnamefont {C.~I.}\ \bibnamefont {Hiley}}, \bibinfo
  {author} {\bibfnamefont {G.~F.~S.}\ \bibnamefont {Whitehead}}, \bibinfo
  {author} {\bibfnamefont {S.}~\bibnamefont {Kar}}, \bibinfo {author}
  {\bibfnamefont {A.~Y.}\ \bibnamefont {Ganin}}, \bibinfo {author}
  {\bibfnamefont {D.}~\bibnamefont {Antypov}}, \bibinfo {author} {\bibfnamefont
  {C.}~\bibnamefont {Collins}}, \bibinfo {author} {\bibfnamefont {M.~S.}\
  \bibnamefont {Dyer}}, \bibinfo {author} {\bibfnamefont {G.}~\bibnamefont
  {Klupp}}, \bibinfo {author} {\bibfnamefont {R.~H.}\ \bibnamefont {Colman}},
  \bibinfo {author} {\bibfnamefont {K.}~\bibnamefont {Prassides}}, \ and\
  \bibinfo {author} {\bibfnamefont {M.~J.}\ \bibnamefont {Rosseinsky}},\ }\href
  {\doibase 10.1038/nchem.2765} {\bibfield  {journal} {\bibinfo  {journal}
  {Nature Chemistry}\ }\textbf {\bibinfo {volume} {9}},\ \bibinfo {pages} {644}
  (\bibinfo {year} {2017})}\BibitemShut {NoStop}%
\bibitem [{\citenamefont {Takabayashi}\ \emph {et~al.}(2017)\citenamefont
  {Takabayashi}, \citenamefont {Menelaou}, \citenamefont {Tamura},
  \citenamefont {Takemori}, \citenamefont {Koretsune}, \citenamefont
  {{\v{S}}tefan{\v{c}}i{\v{c}}}, \citenamefont {Klupp}, \citenamefont {Buurma},
  \citenamefont {Nomura}, \citenamefont {Arita}, \citenamefont {Ar{\v{c}}on},
  \citenamefont {Rosseinsky},\ and\ \citenamefont
  {Prassides}}]{Takabayashi2017}%
  \BibitemOpen
  \bibfield  {author} {\bibinfo {author} {\bibfnamefont {Y.}~\bibnamefont
  {Takabayashi}}, \bibinfo {author} {\bibfnamefont {M.}~\bibnamefont
  {Menelaou}}, \bibinfo {author} {\bibfnamefont {H.}~\bibnamefont {Tamura}},
  \bibinfo {author} {\bibfnamefont {N.}~\bibnamefont {Takemori}}, \bibinfo
  {author} {\bibfnamefont {T.}~\bibnamefont {Koretsune}}, \bibinfo {author}
  {\bibfnamefont {A.}~\bibnamefont {{\v{S}}tefan{\v{c}}i{\v{c}}}}, \bibinfo
  {author} {\bibfnamefont {G.}~\bibnamefont {Klupp}}, \bibinfo {author}
  {\bibfnamefont {A.~J.~C.}\ \bibnamefont {Buurma}}, \bibinfo {author}
  {\bibfnamefont {Y.}~\bibnamefont {Nomura}}, \bibinfo {author} {\bibfnamefont
  {R.}~\bibnamefont {Arita}}, \bibinfo {author} {\bibfnamefont
  {D.}~\bibnamefont {Ar{\v{c}}on}}, \bibinfo {author} {\bibfnamefont {M.~J.}\
  \bibnamefont {Rosseinsky}}, \ and\ \bibinfo {author} {\bibfnamefont
  {K.}~\bibnamefont {Prassides}},\ }\href {\doibase 10.1038/nchem.2764}
  {\bibfield  {journal} {\bibinfo  {journal} {Nature Chemistry}\ }\textbf
  {\bibinfo {volume} {9}},\ \bibinfo {pages} {635} (\bibinfo {year}
  {2017})}\BibitemShut {NoStop}%
\bibitem [{\citenamefont {{\v{S}}tefan{\v{c}}i{\v{c}}}\ \emph
  {et~al.}(2017)\citenamefont {{\v{S}}tefan{\v{c}}i{\v{c}}}, \citenamefont
  {Klupp}, \citenamefont {Knafli{\v{c}}}, \citenamefont {Yufit}, \citenamefont
  {Tav{\v{c}}ar}, \citenamefont {Poto{\v{c}}nik}, \citenamefont {Beeby},\ and\
  \citenamefont {Ar{\v{c}}on}}]{Stefancic2017}%
  \BibitemOpen
  \bibfield  {author} {\bibinfo {author} {\bibfnamefont {A.}~\bibnamefont
  {{\v{S}}tefan{\v{c}}i{\v{c}}}}, \bibinfo {author} {\bibfnamefont
  {G.}~\bibnamefont {Klupp}}, \bibinfo {author} {\bibfnamefont
  {T.}~\bibnamefont {Knafli{\v{c}}}}, \bibinfo {author} {\bibfnamefont {D.~S.}\
  \bibnamefont {Yufit}}, \bibinfo {author} {\bibfnamefont {G.}~\bibnamefont
  {Tav{\v{c}}ar}}, \bibinfo {author} {\bibfnamefont {A.}~\bibnamefont
  {Poto{\v{c}}nik}}, \bibinfo {author} {\bibfnamefont {A.}~\bibnamefont
  {Beeby}}, \ and\ \bibinfo {author} {\bibfnamefont {D.}~\bibnamefont
  {Ar{\v{c}}on}},\ }\href {\doibase 10.1021/acs.jpcc.7b02763} {\bibfield
  {journal} {\bibinfo  {journal} {The Journal of Physical Chemistry C}\
  }\textbf {\bibinfo {volume} {121}},\ \bibinfo {pages} {14864} (\bibinfo
  {year} {2017})}\BibitemShut {NoStop}%
\bibitem [{\citenamefont {Heguri}\ \emph {et~al.}(2015)\citenamefont {Heguri},
  \citenamefont {Kobayashi},\ and\ \citenamefont {Tanigaki}}]{Heguri2015}%
  \BibitemOpen
  \bibfield  {author} {\bibinfo {author} {\bibfnamefont {S.}~\bibnamefont
  {Heguri}}, \bibinfo {author} {\bibfnamefont {M.}~\bibnamefont {Kobayashi}}, \
  and\ \bibinfo {author} {\bibfnamefont {K.}~\bibnamefont {Tanigaki}},\ }\href
  {\doibase 10.1103/PhysRevB.92.014502} {\bibfield  {journal} {\bibinfo
  {journal} {Physical Review B}\ }\textbf {\bibinfo {volume} {92}},\ \bibinfo
  {pages} {014502} (\bibinfo {year} {2015})}\BibitemShut {NoStop}%
\bibitem [{\citenamefont {Kosugi}\ \emph {et~al.}(2011)\citenamefont {Kosugi},
  \citenamefont {Miyake}, \citenamefont {Ishibashi}, \citenamefont {Arita},\
  and\ \citenamefont {Aoki}}]{Kosugi2011}%
  \BibitemOpen
  \bibfield  {author} {\bibinfo {author} {\bibfnamefont {T.}~\bibnamefont
  {Kosugi}}, \bibinfo {author} {\bibfnamefont {T.}~\bibnamefont {Miyake}},
  \bibinfo {author} {\bibfnamefont {S.}~\bibnamefont {Ishibashi}}, \bibinfo
  {author} {\bibfnamefont {R.}~\bibnamefont {Arita}}, \ and\ \bibinfo {author}
  {\bibfnamefont {H.}~\bibnamefont {Aoki}},\ }\href {\doibase
  10.1103/PhysRevB.84.214506} {\bibfield  {journal} {\bibinfo  {journal}
  {Physical Review B}\ }\textbf {\bibinfo {volume} {84}},\ \bibinfo {pages}
  {214506} (\bibinfo {year} {2011})}\BibitemShut {NoStop}%
\bibitem [{\citenamefont {de~Andres}\ \emph
  {et~al.}(2011{\natexlab{a}})\citenamefont {de~Andres}, \citenamefont
  {Guijarro},\ and\ \citenamefont {Verg{\'{e}}s}}]{DeAndres2011a}%
  \BibitemOpen
  \bibfield  {author} {\bibinfo {author} {\bibfnamefont {P.~L.}\ \bibnamefont
  {de~Andres}}, \bibinfo {author} {\bibfnamefont {A.}~\bibnamefont {Guijarro}},
  \ and\ \bibinfo {author} {\bibfnamefont {J.~A.}\ \bibnamefont
  {Verg{\'{e}}s}},\ }\href {\doibase 10.1103/PhysRevB.83.245113} {\bibfield
  {journal} {\bibinfo  {journal} {Physical Review B}\ }\textbf {\bibinfo
  {volume} {83}},\ \bibinfo {pages} {245113} (\bibinfo {year}
  {2011}{\natexlab{a}})}\BibitemShut {NoStop}%
\bibitem [{\citenamefont {de~Andres}\ \emph
  {et~al.}(2011{\natexlab{b}})\citenamefont {de~Andres}, \citenamefont
  {Guijarro},\ and\ \citenamefont {Verg{\'{e}}s}}]{DeAndres2011}%
  \BibitemOpen
  \bibfield  {author} {\bibinfo {author} {\bibfnamefont {P.~L.}\ \bibnamefont
  {de~Andres}}, \bibinfo {author} {\bibfnamefont {A.}~\bibnamefont {Guijarro}},
  \ and\ \bibinfo {author} {\bibfnamefont {J.~A.}\ \bibnamefont
  {Verg{\'{e}}s}},\ }\href {\doibase 10.1103/PhysRevB.84.144501} {\bibfield
  {journal} {\bibinfo  {journal} {Physical Review B}\ }\textbf {\bibinfo
  {volume} {84}},\ \bibinfo {pages} {144501} (\bibinfo {year}
  {2011}{\natexlab{b}})}\BibitemShut {NoStop}%
\bibitem [{\citenamefont {Guijarro}\ and\ \citenamefont
  {Verg{\'{e}}s}(2017)}]{Guijarro2017}%
  \BibitemOpen
  \bibfield  {author} {\bibinfo {author} {\bibfnamefont {A.}~\bibnamefont
  {Guijarro}}\ and\ \bibinfo {author} {\bibfnamefont {J.~A.}\ \bibnamefont
  {Verg{\'{e}}s}},\ }\href {\doibase 10.1103/PhysRevB.95.134112} {\bibfield
  {journal} {\bibinfo  {journal} {Physical Review B}\ }\textbf {\bibinfo
  {volume} {95}},\ \bibinfo {pages} {134112} (\bibinfo {year} {2017})},\
  \Eprint {http://arxiv.org/abs/1702.04141} {arXiv:1702.04141} \BibitemShut
  {NoStop}%
\bibitem [{\citenamefont {Naghavi}\ and\ \citenamefont
  {Tosatti}(2014)}]{Naghavi2014}%
  \BibitemOpen
  \bibfield  {author} {\bibinfo {author} {\bibfnamefont {S.~S.}\ \bibnamefont
  {Naghavi}}\ and\ \bibinfo {author} {\bibfnamefont {E.}~\bibnamefont
  {Tosatti}},\ }\href {\doibase 10.1103/PhysRevB.90.075143} {\bibfield
  {journal} {\bibinfo  {journal} {Physical Review B}\ }\textbf {\bibinfo
  {volume} {90}},\ \bibinfo {pages} {075143} (\bibinfo {year}
  {2014})}\BibitemShut {NoStop}%
\bibitem [{\citenamefont {Wang}\ \emph
  {et~al.}(2011{\natexlab{b}})\citenamefont {Wang}, \citenamefont {Yan},
  \citenamefont {Gui}, \citenamefont {Liu}, \citenamefont {Ying}, \citenamefont
  {Luo},\ and\ \citenamefont {Chen}}]{Wang2011a}%
  \BibitemOpen
  \bibfield  {author} {\bibinfo {author} {\bibfnamefont {X.~F.}\ \bibnamefont
  {Wang}}, \bibinfo {author} {\bibfnamefont {Y.~J.}\ \bibnamefont {Yan}},
  \bibinfo {author} {\bibfnamefont {Z.}~\bibnamefont {Gui}}, \bibinfo {author}
  {\bibfnamefont {R.~H.}\ \bibnamefont {Liu}}, \bibinfo {author} {\bibfnamefont
  {J.~J.}\ \bibnamefont {Ying}}, \bibinfo {author} {\bibfnamefont {X.~G.}\
  \bibnamefont {Luo}}, \ and\ \bibinfo {author} {\bibfnamefont {X.~H.}\
  \bibnamefont {Chen}},\ }\href {\doibase 10.1103/PhysRevB.84.214523}
  {\bibfield  {journal} {\bibinfo  {journal} {Physical Review B}\ }\textbf
  {\bibinfo {volume} {84}},\ \bibinfo {pages} {214523} (\bibinfo {year}
  {2011}{\natexlab{b}})}\BibitemShut {NoStop}%
\bibitem [{\citenamefont {Huang}\ \emph {et~al.}(2014)\citenamefont {Huang},
  \citenamefont {Zhong}, \citenamefont {Zhang}, \citenamefont {Zhao},
  \citenamefont {Zhang}, \citenamefont {Lin},\ and\ \citenamefont
  {Chen}}]{Huang2014}%
  \BibitemOpen
  \bibfield  {author} {\bibinfo {author} {\bibfnamefont {Q.-W.}\ \bibnamefont
  {Huang}}, \bibinfo {author} {\bibfnamefont {G.-H.}\ \bibnamefont {Zhong}},
  \bibinfo {author} {\bibfnamefont {J.}~\bibnamefont {Zhang}}, \bibinfo
  {author} {\bibfnamefont {X.-M.}\ \bibnamefont {Zhao}}, \bibinfo {author}
  {\bibfnamefont {C.}~\bibnamefont {Zhang}}, \bibinfo {author} {\bibfnamefont
  {H.-Q.}\ \bibnamefont {Lin}}, \ and\ \bibinfo {author} {\bibfnamefont
  {X.-J.}\ \bibnamefont {Chen}},\ }\href {\doibase 10.1063/1.4868437}
  {\bibfield  {journal} {\bibinfo  {journal} {The Journal of Chemical Physics}\
  }\textbf {\bibinfo {volume} {140}},\ \bibinfo {pages} {114301} (\bibinfo
  {year} {2014})}\BibitemShut {NoStop}%
\bibitem [{\citenamefont {Dutta}\ and\ \citenamefont
  {Mazumdar}(2014)}]{Dutta2014}%
  \BibitemOpen
  \bibfield  {author} {\bibinfo {author} {\bibfnamefont {T.}~\bibnamefont
  {Dutta}}\ and\ \bibinfo {author} {\bibfnamefont {S.}~\bibnamefont
  {Mazumdar}},\ }\href {\doibase 10.1103/PhysRevB.89.245129} {\bibfield
  {journal} {\bibinfo  {journal} {Physical Review B}\ }\textbf {\bibinfo
  {volume} {89}},\ \bibinfo {pages} {245129} (\bibinfo {year}
  {2014})}\BibitemShut {NoStop}%
\bibitem [{\citenamefont {Yan}\ \emph {et~al.}(2016{\natexlab{a}})\citenamefont
  {Yan}, \citenamefont {Zhang}, \citenamefont {Zhong}, \citenamefont {Ma},\
  and\ \citenamefont {Gao}}]{Yan2016}%
  \BibitemOpen
  \bibfield  {author} {\bibinfo {author} {\bibfnamefont {X.-W.}\ \bibnamefont
  {Yan}}, \bibinfo {author} {\bibfnamefont {C.}~\bibnamefont {Zhang}}, \bibinfo
  {author} {\bibfnamefont {G.}~\bibnamefont {Zhong}}, \bibinfo {author}
  {\bibfnamefont {D.}~\bibnamefont {Ma}}, \ and\ \bibinfo {author}
  {\bibfnamefont {M.}~\bibnamefont {Gao}},\ }\href {\doibase
  10.1039/C6TC04451D} {\bibfield  {journal} {\bibinfo  {journal} {J. Mater.
  Chem. C}\ }\textbf {\bibinfo {volume} {4}},\ \bibinfo {pages} {11566}
  (\bibinfo {year} {2016}{\natexlab{a}})}\BibitemShut {NoStop}%
\bibitem [{\citenamefont {Bl{\"{o}}chl}(1994)}]{PhysRevB.50.17953}%
  \BibitemOpen
  \bibfield  {author} {\bibinfo {author} {\bibfnamefont {P.~E.}\ \bibnamefont
  {Bl{\"{o}}chl}},\ }\href {\doibase 10.1103/PhysRevB.50.17953} {\bibfield
  {journal} {\bibinfo  {journal} {Physical Review B}\ }\textbf {\bibinfo
  {volume} {50}},\ \bibinfo {pages} {17953} (\bibinfo {year}
  {1994})}\BibitemShut {NoStop}%
\bibitem [{\citenamefont {Kresse}\ and\ \citenamefont
  {Hafner}(1993)}]{PhysRevB.47.558}%
  \BibitemOpen
  \bibfield  {author} {\bibinfo {author} {\bibfnamefont {G.}~\bibnamefont
  {Kresse}}\ and\ \bibinfo {author} {\bibfnamefont {J.}~\bibnamefont
  {Hafner}},\ }\href {\doibase 10.1103/PhysRevB.47.558} {\bibfield  {journal}
  {\bibinfo  {journal} {Physical Review B}\ }\textbf {\bibinfo {volume} {47}},\
  \bibinfo {pages} {558} (\bibinfo {year} {1993})}\BibitemShut {NoStop}%
\bibitem [{\citenamefont {Kresse}\ and\ \citenamefont
  {Furthm{\"{u}}ller}(1996)}]{PhysRevB.54.11169}%
  \BibitemOpen
  \bibfield  {author} {\bibinfo {author} {\bibfnamefont {G.}~\bibnamefont
  {Kresse}}\ and\ \bibinfo {author} {\bibfnamefont {J.}~\bibnamefont
  {Furthm{\"{u}}ller}},\ }\href {\doibase 10.1103/PhysRevB.54.11169} {\bibfield
   {journal} {\bibinfo  {journal} {Physical Review B}\ }\textbf {\bibinfo
  {volume} {54}},\ \bibinfo {pages} {11169} (\bibinfo {year}
  {1996})}\BibitemShut {NoStop}%
\bibitem [{\citenamefont {Perdew}\ \emph {et~al.}(1996)\citenamefont {Perdew},
  \citenamefont {Burke},\ and\ \citenamefont
  {Ernzerhof}}]{PhysRevLett.77.3865}%
  \BibitemOpen
  \bibfield  {author} {\bibinfo {author} {\bibfnamefont {J.~P.}\ \bibnamefont
  {Perdew}}, \bibinfo {author} {\bibfnamefont {K.}~\bibnamefont {Burke}}, \
  and\ \bibinfo {author} {\bibfnamefont {M.}~\bibnamefont {Ernzerhof}},\ }\href
  {\doibase 10.1103/PhysRevLett.77.3865} {\bibfield  {journal} {\bibinfo
  {journal} {Physical Review Letters}\ }\textbf {\bibinfo {volume} {77}},\
  \bibinfo {pages} {3865} (\bibinfo {year} {1996})}\BibitemShut {NoStop}%
\bibitem [{\citenamefont {Dion}\ \emph {et~al.}(2004)\citenamefont {Dion},
  \citenamefont {Rydberg}, \citenamefont {Schr{\"{o}}der}, \citenamefont
  {Langreth},\ and\ \citenamefont {Lundqvist}}]{PhysRevLett.92.246401}%
  \BibitemOpen
  \bibfield  {author} {\bibinfo {author} {\bibfnamefont {M.}~\bibnamefont
  {Dion}}, \bibinfo {author} {\bibfnamefont {H.}~\bibnamefont {Rydberg}},
  \bibinfo {author} {\bibfnamefont {E.}~\bibnamefont {Schr{\"{o}}der}},
  \bibinfo {author} {\bibfnamefont {D.~C.}\ \bibnamefont {Langreth}}, \ and\
  \bibinfo {author} {\bibfnamefont {B.~I.}\ \bibnamefont {Lundqvist}},\ }\href
  {\doibase 10.1103/PhysRevLett.92.246401} {\bibfield  {journal} {\bibinfo
  {journal} {Phys. Rev. Lett.}\ }\textbf {\bibinfo {volume} {92}},\ \bibinfo
  {pages} {246401} (\bibinfo {year} {2004})}\BibitemShut {NoStop}%
\bibitem [{\citenamefont {Rom{\'{a}}n-P{\'{e}}rez}\ and\ \citenamefont
  {Soler}(2009)}]{Roman-Perez2009}%
  \BibitemOpen
  \bibfield  {author} {\bibinfo {author} {\bibfnamefont {G.}~\bibnamefont
  {Rom{\'{a}}n-P{\'{e}}rez}}\ and\ \bibinfo {author} {\bibfnamefont {J.~M.}\
  \bibnamefont {Soler}},\ }\href {\doibase 10.1103/PhysRevLett.103.096102}
  {\bibfield  {journal} {\bibinfo  {journal} {Physical Review Letters}\
  }\textbf {\bibinfo {volume} {103}},\ \bibinfo {pages} {096102} (\bibinfo
  {year} {2009})}\BibitemShut {NoStop}%
\bibitem [{\citenamefont {Lee}\ \emph {et~al.}(2010)\citenamefont {Lee},
  \citenamefont {Murray}, \citenamefont {Kong}, \citenamefont {Lundqvist},\
  and\ \citenamefont {Langreth}}]{Lee2010}%
  \BibitemOpen
  \bibfield  {author} {\bibinfo {author} {\bibfnamefont {K.}~\bibnamefont
  {Lee}}, \bibinfo {author} {\bibfnamefont {{\'{E}}.~D.}\ \bibnamefont
  {Murray}}, \bibinfo {author} {\bibfnamefont {L.}~\bibnamefont {Kong}},
  \bibinfo {author} {\bibfnamefont {B.~I.}\ \bibnamefont {Lundqvist}}, \ and\
  \bibinfo {author} {\bibfnamefont {D.~C.}\ \bibnamefont {Langreth}},\ }\href
  {\doibase 10.1103/PhysRevB.82.081101} {\bibfield  {journal} {\bibinfo
  {journal} {Phys. Rev. B}\ }\textbf {\bibinfo {volume} {82}},\ \bibinfo
  {pages} {081101} (\bibinfo {year} {2010})}\BibitemShut {NoStop}%
\bibitem [{\citenamefont {Klime{\v{s}}}\ \emph {et~al.}(2011)\citenamefont
  {Klime{\v{s}}}, \citenamefont {Bowler},\ and\ \citenamefont
  {Michaelides}}]{Klimes2011}%
  \BibitemOpen
  \bibfield  {author} {\bibinfo {author} {\bibfnamefont {J.}~\bibnamefont
  {Klime{\v{s}}}}, \bibinfo {author} {\bibfnamefont {D.~R.}\ \bibnamefont
  {Bowler}}, \ and\ \bibinfo {author} {\bibfnamefont {A.}~\bibnamefont
  {Michaelides}},\ }\href {\doibase 10.1103/PhysRevB.83.195131} {\bibfield
  {journal} {\bibinfo  {journal} {Phys. Rev. B}\ }\textbf {\bibinfo {volume}
  {83}},\ \bibinfo {pages} {195131} (\bibinfo {year} {2011})},\ \Eprint
  {http://arxiv.org/abs/arXiv:1102.1358v1} {arXiv:arXiv:1102.1358v1}
  \BibitemShut {NoStop}%
\bibitem [{\citenamefont {Yan}\ \emph {et~al.}(2016{\natexlab{b}})\citenamefont
  {Yan}, \citenamefont {Wang}, \citenamefont {Gao}, \citenamefont {Ma},\ and\
  \citenamefont {Huang}}]{Yan2016a}%
  \BibitemOpen
  \bibfield  {author} {\bibinfo {author} {\bibfnamefont {X.-W.}\ \bibnamefont
  {Yan}}, \bibinfo {author} {\bibfnamefont {Y.}~\bibnamefont {Wang}}, \bibinfo
  {author} {\bibfnamefont {M.}~\bibnamefont {Gao}}, \bibinfo {author}
  {\bibfnamefont {D.}~\bibnamefont {Ma}}, \ and\ \bibinfo {author}
  {\bibfnamefont {Z.}~\bibnamefont {Huang}},\ }\href {\doibase
  10.1021/acs.jpcc.6b08373} {\bibfield  {journal} {\bibinfo  {journal} {The
  Journal of Physical Chemistry C}\ }\textbf {\bibinfo {volume} {120}},\
  \bibinfo {pages} {22565} (\bibinfo {year} {2016}{\natexlab{b}})}\BibitemShut
  {NoStop}%
\bibitem [{\citenamefont {Zhong}\ \emph {et~al.}(2017)\citenamefont {Zhong},
  \citenamefont {Zhang}, \citenamefont {Yan}, \citenamefont {Li}, \citenamefont
  {Du}, \citenamefont {Jing},\ and\ \citenamefont {Ma}}]{Zhong2017}%
  \BibitemOpen
  \bibfield  {author} {\bibinfo {author} {\bibfnamefont {G.-H.}\ \bibnamefont
  {Zhong}}, \bibinfo {author} {\bibfnamefont {C.}~\bibnamefont {Zhang}},
  \bibinfo {author} {\bibfnamefont {X.}~\bibnamefont {Yan}}, \bibinfo {author}
  {\bibfnamefont {X.}~\bibnamefont {Li}}, \bibinfo {author} {\bibfnamefont
  {Z.}~\bibnamefont {Du}}, \bibinfo {author} {\bibfnamefont {G.}~\bibnamefont
  {Jing}}, \ and\ \bibinfo {author} {\bibfnamefont {C.}~\bibnamefont {Ma}},\
  }\href {\doibase 10.1080/00268976.2016.1274439} {\bibfield  {journal}
  {\bibinfo  {journal} {Molecular Physics}\ }\textbf {\bibinfo {volume}
  {115}},\ \bibinfo {pages} {472} (\bibinfo {year} {2017})}\BibitemShut
  {NoStop}%
\bibitem [{\citenamefont {Wang}\ \emph {et~al.}(2014)\citenamefont {Wang},
  \citenamefont {Selbach},\ and\ \citenamefont {Grande}}]{Wang2014a}%
  \BibitemOpen
  \bibfield  {author} {\bibinfo {author} {\bibfnamefont {Z.}~\bibnamefont
  {Wang}}, \bibinfo {author} {\bibfnamefont {S.~M.}\ \bibnamefont {Selbach}}, \
  and\ \bibinfo {author} {\bibfnamefont {T.}~\bibnamefont {Grande}},\ }\href
  {\doibase 10.1039/c3ra47187j} {\bibfield  {journal} {\bibinfo  {journal} {RSC
  Advances}\ }\textbf {\bibinfo {volume} {4}},\ \bibinfo {pages} {4069}
  (\bibinfo {year} {2014})}\BibitemShut {NoStop}%
\bibitem [{\citenamefont {Henkelman}\ \emph {et~al.}(2006)\citenamefont
  {Henkelman}, \citenamefont {Arnaldsson},\ and\ \citenamefont
  {J{\'{o}}nsson}}]{Henkelman2006}%
  \BibitemOpen
  \bibfield  {author} {\bibinfo {author} {\bibfnamefont {G.}~\bibnamefont
  {Henkelman}}, \bibinfo {author} {\bibfnamefont {A.}~\bibnamefont
  {Arnaldsson}}, \ and\ \bibinfo {author} {\bibfnamefont {H.}~\bibnamefont
  {J{\'{o}}nsson}},\ }\href {\doibase 10.1016/j.commatsci.2005.04.010}
  {\bibfield  {journal} {\bibinfo  {journal} {Computational Materials Science}\
  }\textbf {\bibinfo {volume} {36}},\ \bibinfo {pages} {354} (\bibinfo {year}
  {2006})}\BibitemShut {NoStop}%
\bibitem [{\citenamefont {Wang}\ \emph {et~al.}(2012)\citenamefont {Wang},
  \citenamefont {Luo}, \citenamefont {Ying}, \citenamefont {Xiang},
  \citenamefont {Zhang}, \citenamefont {Zhang}, \citenamefont {Zhang},
  \citenamefont {Yan}, \citenamefont {Wang}, \citenamefont {Cheng},
  \citenamefont {Ye},\ and\ \citenamefont {Chen}}]{Wang2012}%
  \BibitemOpen
  \bibfield  {author} {\bibinfo {author} {\bibfnamefont {X.~F.}\ \bibnamefont
  {Wang}}, \bibinfo {author} {\bibfnamefont {X.~G.}\ \bibnamefont {Luo}},
  \bibinfo {author} {\bibfnamefont {J.~J.}\ \bibnamefont {Ying}}, \bibinfo
  {author} {\bibfnamefont {Z.~J.}\ \bibnamefont {Xiang}}, \bibinfo {author}
  {\bibfnamefont {S.~L.}\ \bibnamefont {Zhang}}, \bibinfo {author}
  {\bibfnamefont {R.~R.}\ \bibnamefont {Zhang}}, \bibinfo {author}
  {\bibfnamefont {Y.~H.}\ \bibnamefont {Zhang}}, \bibinfo {author}
  {\bibfnamefont {Y.~J.}\ \bibnamefont {Yan}}, \bibinfo {author} {\bibfnamefont
  {a.~F.}\ \bibnamefont {Wang}}, \bibinfo {author} {\bibfnamefont
  {P.}~\bibnamefont {Cheng}}, \bibinfo {author} {\bibfnamefont {G.~J.}\
  \bibnamefont {Ye}}, \ and\ \bibinfo {author} {\bibfnamefont {X.~H.}\
  \bibnamefont {Chen}},\ }\href {\doibase 10.1088/0953-8984/24/34/345701}
  {\bibfield  {journal} {\bibinfo  {journal} {Journal of physics: Condensed
  matter}\ }\textbf {\bibinfo {volume} {24}},\ \bibinfo {pages} {345701}
  (\bibinfo {year} {2012})}\BibitemShut {NoStop}%
\bibitem [{\citenamefont {Phan}\ \emph {et~al.}(2014)\citenamefont {Phan},
  \citenamefont {Heguri}, \citenamefont {Tanabe}, \citenamefont {Shimotani},
  \citenamefont {Nakano}, \citenamefont {Nozue},\ and\ \citenamefont
  {Tanigaki}}]{Phan2014}%
  \BibitemOpen
  \bibfield  {author} {\bibinfo {author} {\bibfnamefont {Q.~T.~N.}\
  \bibnamefont {Phan}}, \bibinfo {author} {\bibfnamefont {S.}~\bibnamefont
  {Heguri}}, \bibinfo {author} {\bibfnamefont {Y.}~\bibnamefont {Tanabe}},
  \bibinfo {author} {\bibfnamefont {H.}~\bibnamefont {Shimotani}}, \bibinfo
  {author} {\bibfnamefont {T.}~\bibnamefont {Nakano}}, \bibinfo {author}
  {\bibfnamefont {Y.}~\bibnamefont {Nozue}}, \ and\ \bibinfo {author}
  {\bibfnamefont {K.}~\bibnamefont {Tanigaki}},\ }\href {\doibase
  10.1039/c4dt00071d} {\bibfield  {journal} {\bibinfo  {journal} {Dalton
  Transactions}\ }\textbf {\bibinfo {volume} {43}},\ \bibinfo {pages} {10040}
  (\bibinfo {year} {2014})}\BibitemShut {NoStop}%
\bibitem [{\citenamefont {Zhong}\ \emph {et~al.}(2014)\citenamefont {Zhong},
  \citenamefont {Huang},\ and\ \citenamefont {Lin}}]{Zhong2014}%
  \BibitemOpen
  \bibfield  {author} {\bibinfo {author} {\bibfnamefont {G.}~\bibnamefont
  {Zhong}}, \bibinfo {author} {\bibfnamefont {Z.}~\bibnamefont {Huang}}, \ and\
  \bibinfo {author} {\bibfnamefont {H.}~\bibnamefont {Lin}},\ }\href {\doibase
  10.1109/TMAG.2014.2329602} {\bibfield  {journal} {\bibinfo  {journal} {IEEE
  Transactions on Magnetics}\ }\textbf {\bibinfo {volume} {50}},\ \bibinfo
  {pages} {1700103} (\bibinfo {year} {2014})}\BibitemShut {NoStop}%
\end{thebibliography}%

\end{document}